\documentstyle[amssymb,aps,preprint]{revtex}

\tightenlines

\begin{document}
\title{Theory of melting of vortex lattice in high $T_{c}$ superconductors}
\author{Dingping Li\thanks{%
e-mail: lidp@mono1.math.nctu.edu.tw} and Baruch Rosenstein\thanks{%
e-mail: baruch@vortex1.ep.nctu.edu.tw}}
\address{{\it National Center for Theoretical Sciences and} \\
{\it Electrophysics Department, National Chiao Tung University } \\
{\it Hsinchu 30050, Taiwan, R. O. C.}}
\date{\today}
\maketitle

\begin{abstract}
Theory of melting of the vortex lattice in type II superconductors in the
framework of Ginzburg - Landau approach is presented. The melting line
location is determined and magnetization and specific heat jumps along it
are calculated . The magnetization of liquid is larger than that of solid by 
$1.8\%$ irrespective of the melting temperature, while the specific heat
jump is about $6\%$ and decreases slowly with temperature. The magnetization
curves agrees with experimental results on $YBCO$ and Monte Carlo
simulations.
\end{abstract}


\newpage

\section{Introduction and the main idea}

Abrikosov vortices created by magnetic field in type II superconductors
strongly interact with each other creating highly correlated configurations
like the vortex lattice. In high $T_{c}$ cuprates thermal fluctuations at
relatively large temperatures are strong enough to melt the lattice. Several
recent remarkable experiments discovered that the vortex lattice melting in
high $T_{c}$ superconductors is first order with magnetization jumps \cite
{Zeldov} and spikes in specific heat \cite{Schilling}. Magnetization and
entropy jumps were measured using great variety of techniques: local Hall
probes \cite{Zeldov}, SQUID\cite{Pastoriza,Welp}, torque magnetometry \cite
{Willemin,Nishizaki} and integrating the specific heat spike \cite{Schilling}%
. It was found that in addition to the spike there is also a jump in
specific heat which was measured as well \cite{Schilling,Roulin}. These
precise measurements pose a question of accurate quantitative theoretical
description of thermal fluctuations in vortex matter.

The Ginzburg - Landau (GL) approach is appropriate to describe thermal
fluctuations near $T_{c}$. The GL model is however highly nontrivial even
within the lowest Landau level (LLL) approximation valid at high fields. In
this simplified model the only parameter is the dimensionless scaled
temperature $a_{T}$ $\thicksim $ $(T-T_{mf}(H))/(TH)^{2/3}$ (defined more
precisely in eq.(\ref{athl}) below). Over last twenty years great variety of
theoretical methods were applied to study this model. Brezin, Nelson and
Thiaville \cite{Brezin} applied the renormalization group (RG) on the one
loop level. No fixed points of the (functional) RG equations were found and
they concluded therefore that the transition from liquid to the solid is
first order \cite{MooreRG}. This approach however does not provide a
quantitative theory of melting.

Two perturbative approaches were developed and greatly improved recently to
describe the solid phase and the liquid phase respectively.The perturbative
approach on the liquid side was pioneered by Thouless and Ruggeri \cite
{Ruggeri}. They developed a perturbative expansion around a homogeneous
(liquid) state in which all the ''bubble'' diagrams are resummed.
Unfortunately they found that the series are asymptotic and although first
few terms provide accurate results at very high temperatures, the series
become inapplicable for $a_{T}$ less than $-2$ which is quite far above the
melting line (believed to be located around $a_{T}\sim -10$). We recently
obtained the optimized gaussian series \cite{LiPRL} which are convergent
rather than asymptotic with radius of convergence of $a_{T}=-5$ ($a_{T}=-5$
is still above the melting point).

On the solid side, Eilenberger and Maki and Takayama \cite{Eilenberger}
calculated the fluctuations spectrum around Abrikosov's mean field solution.
They noticed that the vortex lattice phonon modes are softer than that of
the acoustic phonons in atomic crystals and this leads to infrared
divergences in certain quantities. This was initially interpreted as
destruction of the vortex solid by thermal fluctuations and the perturbation
theory was abandoned. However the divergence looks suspiciously similar to
''spurious'' IR divergences in the critical phenomena theory and recently it
was shown that all these IR divergences cancel in physical quantities \cite
{Rosenstein}. The series therefore are reliable, and were extended to two
loops, so that the LLL GL theory on the solid side is now precise enough
even at melting point.

However on the liquid side one needs a theory in the region $-10<a_{T}<-5$.
Moreover this theory should be very precise since free energies of solid and
liquid differ only by few percents near melting. This requires a better
qualitative understanding of the metastable phases of the theory. It is
clear that the overheated solid becomes unstable at some finite temperature.
It not clear however whether overcooled liquid becomes unstable at some
finite temperature (like water) or exists all the way down to $T=0$ as a
metastable state. The gaussian (Hartree - Fock) variational calculation,
although perhaps of a limited precision, is usually a very good guide as far
as qualitative features of the phase diagram are concerned. Such a
calculation in the liquid was performed long ago \cite{Ruggeri}, while more
complicated one sampling also inhomogeneous states (vortex lattice) was
obtained recently \cite{LiPRL}. The gaussian results are as follows. The
free energy of the solid state is lower than that of the liquid for
temperatures lower than melting temperature. The solid state is therefore
the stable one below it, becomes metastable at somewhat higher temperatures
and is destabilized at $a_{T}=-5$. The liquid state becomes metastable below
the melting temperature, but unlike the solid, does not loose metastability
all the way down to $T=0$. The excitation energy of the overcooled liquid
approaches zero as a power $\varepsilon \sim 1/a_{T}^{2}$.

Meantime in different area of physics similar qualitative results were
obtained. It was shown by variety of analytical and numerical methods that
liquid (gas) phase of the classical one component Coulomb plasma exists as a
metastable state down to $T=0$ with energy gradually approaching that of the
Madelung solid and excitation energy diminishing \cite{Carvalho}. It seems
plausible to speculate that the same would happen in any system of particles
interacting via long range repulsive forces. In fact the vortices in the
London approximation become a sort of repelling particles with the force
even more long range than Coulombic. This was an additional strong
motivation to consider the above scenario in vortex matter. Below we provide
both theoretical and phenomenological evidence that the above scenario is
the correct one.

Assuming absence of singularities on the liquid branch allows to develop an
essentially precise theory of the LLL GL model in vortex liquid (even
including overcooled liquid) using the Borel - Pade (BP) \cite{Baker} method
at any temperature. First we clarify several issues which prevented the use
and acceptance of the BP method in the past and then combine it with the
recently developed LLL theory of solids to calculate the melting line and
magnetization and specific heat jumps. Very early Ruggeri and Thouless \cite
{Ruggeri} tried to use BP to calculate the specific heat without much
success because their series were too short \cite{Wilkin}. Attempts to use
BP for the calculation of melting also ran into problems. Hikami, Fujita and
Larkin \cite{Hikami} tried to find the melting point by comparing the BP
energy with the one loop solid energy and obtained $a_{T}=-7$. However their
one loop solid energy was incorrect and in any case it was not precise
enough.

The LLL GL model was also studied numerically in both 3D \cite{Sasik} and 2D 
\cite{MC} and by a variety of nonperturbative analytical methods. Among them
the density functional\cite{Herbut}, $1/N$ \cite{largeN}, dislocation theory
of melting \cite{Maniv} and others \cite{Andreev}.

As we show in this paper, the BP energy combined with the correct two loop
solid energy computed recently gives scaled melting temperature $%
a_{T}^{m}=-9.5$ and in addition predicts other characteristics of the model.
The melting line location is determined and magnetization and specific heat
jumps are calculated . The magnetization of liquid is larger than that of
solid by $1.8\%$ irrespective of the melting temperature, while the specific
heat jump is about $6\%$ and decreases slowly with temperature.

In addition to theory of melting we calculated magnetization and specific
heat curves. The magnetization curves agree quite well with Monte Carlo
simulations of the LLL GL \cite{Sasik}, and almost perfectly for specific
heat in 2D by Kato and Nagaosa in ref.\cite{MC}. However to describe
experimental results like $YBCO$ at lower fields and lower temperature,
higher Landau levels (HLL) corrections are required. Experimentally it was
claimed that one can establish the LLL scaling for fields above $3T$ \cite
{Sok}. A glance at the data however shows that above $T_{c}$ the scaling for
magnetization curves is generally very bad: the magnetization due to the LLL
contribution is much larger that the experimental one above $T_{c}$.
Therefore we calculated the leading correction by \ the HLLs and then
compare with experiments.

The paper is organized as follows. The models are defined in section II and
the melting theory in the LLL is described in section III. In section IV,
the HLL correction is discussed and the magnetization curves are compared
with experiments. We conclude in section V.

\section{The model and basic assumptions}

\subsection{The model}

Thermal fluctuations of 3D materials with not very strong asymmetry along
the $z$ axis are effectively described by the following Ginzburg-Landau free
energy: 
\begin{equation}
F=\int d^{3}x\frac{{\hbar }^{2}}{2m_{ab}}\left| {\bf D}\psi \right| ^{2}+%
\frac{{\hbar }^{2}}{2m_{c}}|\partial _{z}\psi |^{2}-a(T)|\psi |^{2}+\frac{%
b^{\prime }}{2}|\psi |^{4}+\frac{\left( {\bf B}-{\bf H}\right) ^{2}}{8\pi }
\label{GLdef}
\end{equation}
where ${\bf A}=(By,0)$ describes magnetic field (considered constant and
nonfluctuating, see below) in Landau gauge and covariant derivative is
defined by ${\bf D}\equiv {\bf \nabla }-i\frac{2\pi }{\Phi _{0}}{\bf A,}\Phi
_{0}\equiv \frac{hc}{e^{\ast }}$($e^{\ast }=2e$). Statistical physics is
described by the statistical sum: 
\begin{equation}
Z=\int D\psi D\overline{\psi }\exp \left( -\frac{F}{T}\right)
\end{equation}
Our aim is to quantitatively describe the effects of thermal fluctuations of
high $T_{c}$ cuprates of the $YBCO$ type on the few percent precision level
(such a precision is required for the theory of melting since energies of
liquid and solid near melting differ only by a few percent).

\subsection{Assumptions}

The use of the above GL energy makes several physical assumptions. They are
listed below.

(1) {\it Continuum model}

We use anisotropic GL model despite the layered structure of the high $T_{c} 
$ cuprates for which the Lawrence - Doniah model is more appropriate model.
Effects of layered structure are dominant in $BSCCO$ or $Tl$ compounds ( $%
\gamma \equiv \sqrt{m_{c}/m_{ab}}>1000$) and noticeable for cuprates with
anisotropy of order $\gamma =50$ like $LaBaCuO$ or $Hg1223$. The
requirement, that the GL can be effectively used, therefore limits us to
optimally doped $YBCO_{7-\delta }$ (or slightly overdoped or underdoped) for
which the anisotropy parameter is not very large $\gamma =4-8$), $DyBCO$ or
possibly $Hg1221$ which has a slightly larger anisotropy.

(2) {\it Range of validity of using the mesoscopic (GL) approach}

The GL approach generally is an effective mesoscopic approach applicable
when we can neglect higher order terms generated when one ''integrates out''
microscopic degrees of freedom typically but not always near second order
phase transitions. The leading higher dimensional terms we neglect (as
''irrelevant'') are $|\psi |^{6}$ and higher (four) derivative terms. After
that the model practically becomes rotationally symmetric in the $ab$ plane.
We can rescale it to $m_{a}\simeq m_{b}=m_{ab}.$ For several physical
questions this assumption is not valid because irrelevant terms neglected
might become ''dangerous''. For example the question of the structural phase
transition into the square lattice is clearly of this type \cite{structural}%
. It is known that even assuming $m_{a}/m_{b}=1$ in low temperature vortex
lattices in $YBCO$ rotational symmetry is broken down to the fourfold
symmetry by the four derivative terms. However there is no significant
correction to the magnetization from the those higher dimension terms.

(3) {\it Expansion of parameters around }$T_{c}$

Generally parameters of the GL model of eq.(\ref{GLdef}) are complicated
functions of temperature which is determined by the details of the
microscopic theory. We expand the coefficient $a(T)$ near $T_{c}:$ 
\begin{equation}
a(T)=T_{c}[\alpha (1-t)-\alpha ^{\prime }(1-t)^{2}+...],
\end{equation}
where $t\equiv T/T_{c}$. The second and higher terms in the expansion are
omitted and therefore when temperature deviates significantly from $T_{c}$
one cannot expect the model to have a good precision.

(4) {\it Constant nonfluctuating magnetic field}

For strongly type II superconductors like the high $T_{c}$ cuprates not very
far from $H_{c2}(T)$ (this easily covers the range of interest in this
paper, for the detailed discussion of the range of applicability beyond it
see ref.\cite{LiHLL}) magnetic field is homogeneous to a high degree due to
superposition from many vortices. Inhomogeneity is of order $1/\kappa
^{2}\sim 10^{-3}$. Since the main subject of this study is the thermal
fluctuations effects of the order parameter field, one might ask whether
thermal fluctuations of the electromagnetic field should be also taken into
account. Halperin, Lubensky and Ma considered this question long ago\cite
{HLM}. The conclusion was that they are completely negligible for very large 
$\kappa .$ Upon discovery of the high $T_{c}$ cuprates, the issue was
reconsidered \cite{Lobb} and the same result was obtained to a very high
precision. Therefore here magnetic field is treated both as constant and
nonfluctuating ($B=H$) and the last term in eq.(\ref{GLdef}) can be omitted
(to precision of order $1/\kappa ^{2}$). However when we calculate \ the
magnetization, $M=(B-H)/4\pi $ which is of order $1/\kappa ^{2}$, high order
correction must be considered.

(5) {\it Disorder.}

Point - like disorder is always present in $YBCO$. For example when the
melting line becomes the second order transition and magnetization becomes
irreversible the \ systems show clear disorder effects. However in some
samples the disorder effects are minor. In the maximally oxidized sample 
\cite{Nishizaki} the second order transition is not seen even at highest
available fields ($30T$). In the optimally doped sample the same is true up
to $12T$ \cite{Paulius}. The situation is believed to be different in other
materials like $DyBa_{2}Cu_{3}O_{7}$ \cite{Roulin,Garfield} or twinned $YBCO$%
. We will address the disorder problem in our future publication.

\subsection{Landau level modes in the quasimomentum basis}

Assuming that all the requirements are met we now divide the fluctuations
into the LLL and HLL modes. Throughout most of the paper will use the
coherence length $\xi =\sqrt{{\hbar }^{2}/\left( 2m_{ab}\alpha T_{c}\right) }
$ as a unit of length, $T_{c}$ as unit of temperature, $T=tT_{c}$, and $%
\frac{dH_{c2}(T_{c})}{dT}T_{c}=\frac{\Phi _{0}}{2\pi \xi ^{2}}$ as a unit of
magnetic field, $B=bH_{c2}$. After rescaling eq.(\ref{GLdef})by $%
x\rightarrow \xi x,y\rightarrow \xi y,z\rightarrow \frac{\xi z}{\gamma }%
,\psi ^{2}\rightarrow \frac{2\alpha T_{c}}{b^{\prime }}\psi ^{2}$ ($\gamma
\equiv \sqrt{m_{c}/m_{ab}}$) one obtains: 
\begin{equation}
f=\frac{F}{T}=\frac{1}{\omega }\int d^{3}x\left[ \frac{1}{2}|{\bf D}\psi
|^{2}+\frac{1}{2}|\partial _{z}\psi |^{2}-(a_{h}+\frac{b}{2})|\psi |^{2}+%
\frac{1}{2}|\psi |^{4}+\frac{\kappa ^{2}\left( {\bf b}-{\bf h}\right) ^{2}}{4%
}\right] ,  \label{GLs1}
\end{equation}
where 
\begin{equation}
\omega =\sqrt{2Gi}\pi ^{2}t;\ \ \ a_{h}=\frac{1-t-b}{2}
\end{equation}
The Ginzburg number is given by 
\begin{equation}
Gi\equiv \frac{1}{2}\left( \frac{32\pi e^{2}\kappa ^{2}\xi T_{c}\gamma }{%
c^{2}h^{2}}\right) ^{2}
\end{equation}
It is convenient to expand the order parameter field in a complete basis of
noninteracting theory: Landau levels. In the hexagonal lattice phase the
most convenient basis is the quasimomentum basis: 
\begin{equation}
\psi (x)=\frac{1}{\sqrt{2}}\int_{k}\frac{e^{-ik_{3}x_{3}}}{\sqrt{2\pi }}%
\sum_{n=0}^{\infty }\frac{\varphi _{{\bf k}}^{n}(x)}{\left( \sqrt{2\pi }%
\right) ^{2}}\psi ^{n}(k).
\end{equation}
Here $\varphi _{{\bf k}}^{n}(x)$ the $n^{th}$ Landau level with
quasi-momentum ${\bf k}$: 
\begin{eqnarray}
\varphi _{{\bf k}}^{n} &=&\sqrt{\frac{2\pi }{\sqrt{\pi }2^{n}n!a}}%
\sum\limits_{l=-\infty }^{\infty }H_{n}(y\sqrt{b}+\frac{k_{x}}{\sqrt{b}}-%
\frac{2\pi }{a}l)  \label{quasim} \\
&&\times \exp \left\{ i\left[ \frac{\pi l(l-1)}{2}+\frac{2\pi (\sqrt{b}x-%
\frac{k_{y}}{\sqrt{b}})}{a}l-xk_{x}\right] -\frac{1}{2}(y\sqrt{b}+\frac{k_{x}%
}{\sqrt{b}}-\frac{2\pi }{a}l)^{2}\right\} .  \nonumber
\end{eqnarray}
Even in the liquid state which is more symmetric than the hexagonal lattice
we find it convenient to use this basis.

\section{Melting line, magnetization and specific heat in LLL approximation}

The LLL contributes significantly the thermal fluctuation which in
particular leads to the melting (for example, ref.(\cite{Brezin}), and
Pierson and Valls in ref. (\cite{Sok}) {\it et al}). In this section, we
will essentially solve the LLL GL model.

\subsection{The LLL scaling}

If the magnetic field is quite high, \ we can keep only the $n=0$ LLL modes
(in eq.(\ref{GLs1}), we use the LLL condition $|{\bf D}\psi |^{2}=b|\psi
|^{2}$ ): 
\begin{equation}
f=\frac{F}{T}=\frac{1}{\omega }\int d^{3}x\left[ \frac{1}{2}|\partial
_{z}\psi |^{2}-a_{h}|\psi |^{2}+\frac{1}{2}|\psi |^{4}+\frac{\kappa
^{2}\left( {\bf b}-{\bf h}\right) ^{2}}{4}\right] .
\end{equation}
Within the LLL approximation the problem simplifies (no gradient term in
directions perpendicular to the field) and possesses an LLL scaling \cite
{Thouless}. After additional rescaling $x\rightarrow x/\sqrt{b},y\rightarrow
y/\sqrt{b},z\rightarrow z\left( \frac{b\omega }{4\pi \sqrt{2}}\right)
^{-1/3},\psi ^{2}\rightarrow \left( \frac{b\omega }{4\pi \sqrt{2}}\right)
^{2/3}\psi ^{2}$, the dimensionless free energy becomes: 
\begin{equation}
f=\frac{1}{4\pi \sqrt{2}}\int d^{3}x\left[ \frac{1}{2}|\partial _{z}\psi
|^{2}+a_{T}|\psi |^{2}+\frac{1}{2}|\psi |^{4}+\left( \frac{b\omega }{4\pi 
\sqrt{2}}\right) ^{-4/3}\frac{\kappa ^{2}\left( {\bf b}-{\bf h}\right) ^{2}}{%
4}\right]   \label{LLL2}
\end{equation}
and it can be approximated to order \ $\frac{1}{\kappa ^{2}}$ 
\begin{equation}
\frac{1}{4\pi \sqrt{2}}\int d^{3}x\left[ \frac{1}{2}|\partial _{z}\psi
|^{2}+a_{T}|\psi |^{2}+\frac{1}{2}|\psi |^{4}\right]   \label{LLL1}
\end{equation}
where the scaled temperature 
\begin{mathletters}
\begin{equation}
a_{T}=-\left( \frac{b\omega }{4\pi \sqrt{2}}\right) ^{-2/3}a_{h}.
\label{athl}
\end{equation}
Free energy density in the newly scaled model is: 
\end{mathletters}
\begin{equation}
f_{eff}=-\frac{4\pi \sqrt{2}}{V}\ln \int D\psi D\overline{\psi }\exp \left(
-f\right) 
\end{equation}
is a function of $a_{T}$ only to order $\frac{1}{\kappa ^{2}}$.

\subsection{Free energy of liquid and solid}

Now we specify the solution of the LLL GL model. The liquid LLL (scaled)
free energy is written as 
\begin{equation}
f_{liq}=4\varepsilon ^{1/2}[1+g\left( x\right) ].
\end{equation}
The function $g$ can be expanded as 
\begin{equation}
g\left( x\right) =\sum c_{n}x^{n},
\end{equation}
where the high temperature small parameter $x=\frac{1}{2}\varepsilon ^{-3/2}$
is defined as a solution of the Gaussian gap equation 
\begin{equation}
\varepsilon ^{3/2}-a_{T}\varepsilon ^{1/2}-4=0
\end{equation}
for the excitation energy $\varepsilon $.\ The coefficients $c_{n}$ can be
found in \cite{Hikami}. We will denote by $g_{k}\left( x\right) $ the $%
[k,k-1]$ BP transform \cite{Baker} of $g(x)$ (other BP approximants clearly
violate the correct low temperature asymptotics).\ The BP transform is
defined as $\int_{0}^{\infty }g_{k}^{\prime }\left( xt\right) \exp \left(
-t\right) dt$ where $g_{k}^{\prime }$ is the $[k,k-1]$ Pade transform of $%
\sum_{n=1}^{2k-1}\frac{c_{n}x^{n}}{n!}$.

\ For $k=4$,and $k=5$, the liquid energy converges to required precision ($%
0.1\%$). In what follows we will use $g_{5}$.

The solid energy to two loops is \cite{Rosenstein,LiSolid}: 
\begin{equation}
f_{sol}=-\frac{a_{T}^{2}}{2\beta _{A}}+2.848\left| a_{T}\right| ^{1/2}+\frac{%
2.4}{a_{T}}.
\end{equation}
where $\beta _{A}=1.1596$. On Fig.1 we plot the energies of solid and
liquid. They are very close (see the difference on inset). The liquid energy
completely agrees with the optimized gaussian expansion results \cite{LiPRL}
till its radius of convergence at $a_{T}=-5$. The energy of the overcooled
liquid at low temperatures approaches that of the vortex solid. We obtained
similar results in 2D and found that they also agree with Monte Carlo
simulations \cite{MC} and optimized gaussian expansion \cite{LiPRL} (see a
brief discussion in the next section)

\subsection{Melting line. Comparison with experiments, Monte Carlo
simulations and Lindemann criterion}

Comparing solid and liquid energy (inset in Fig.1), we find that $%
a_{T}^{m}=-9.5.$ This is in accord with experimental results. As an example
on Fig.2 we present fitting of the melting line of fully oxidized $%
YBa_{2}Cu_{3}O_{7}$ \cite{Nishizaki} which gives $%
T_{c}=88.2,H_{c2}=175.9,Gi=7.0$ $10^{-5}$. Melting lines of optimally doped
untwinned \cite{Schilling,Welp,Willemin} $YBa_{2}Cu_{3}O_{7-\delta }$ and $%
DyBa_{2}Cu_{3}O_{7}$ \cite{Roulin} are also fitted extremely well. The
results of fitting are given in Table 1.

\begin{center}
{\bf Table 1}

Parameters of high $T_{c}$ superconductors deduced from the melting line

\begin{tabular}{|c|c|c|c|c|c|c|}
\hline
material & $T_{c}$ & $H_{c2}$ & $Gi$ & $\kappa $ & $\gamma $ & reference \\ 
\hline
$YBCO_{7-\delta }$ & $93.07$ & $167.53$ & $1.910^{-4}$ & $48.5$ & $7.76$ & 
\cite{Schilling} \\ \hline
$YBCO_{7}$ & $88.16$ & $175.9$ & $7.010^{-5}$ & $50$ & $4$ & \cite{Nishizaki}
\\ \hline
$DyBCO_{6.7}$ & $90.14$ & $163$ & $3.210^{-5}$ & $33.77$ & $5.3$ & \cite
{Roulin} \\ \hline
\end{tabular}
\end{center}

The available 3D Monte Carlo simulations \cite{Sasik} unfortunately are not
precise enough to provide an accurate melting point since the LLL scaling is
violated and one gets values of $a_{T}^{m}=-14.5,-13.2,-10.9$ at magnetic
fields $1,2,5T$ respectively. We found also that the theoretical
magnetization calculated by using parameters given by ref.\cite{Sasik} is in
a very good agreement with the Monte Carlo simulation result of ref.\cite
{Sasik}. However the determination of melting temperature needs higher
precision, and the sample size ($\sim $100 vortices) used in ref.\cite{Sasik}
may be not large enough to give an accurate determination of the melting
temperature (due to boundary effects, LLL scaling will be violated too). The
situation in 2D is better since the sample size is much larger. We performed
similar calculation for the 2D LLL GL liquid free energy, combined it with
the earlier solid energy calculation \cite{Rosenstein,LiSolid} 
\begin{equation}
f_{sol}=-\frac{a_{T}^{2}}{2\beta _{A}}+2\log \frac{\left| a_{T}\right| }{%
4\pi ^{2}}-\frac{19.9}{a_{T}^{2}}-2.92.  \label{perbresl}
\end{equation}
and find that the melting point $a_{T}^{m}=-13.2$. It is \ in good agreement
with MC simulations \cite{MC}.

Phenomenologically melting line can be located using Lindemann criterion or
its more refined version using Debye - Waller factor. The more refined
definition is required since vortices are not pointlike. It was found
numerically for Yukawa gas \cite{Stevens} that the Debye - Waller factor $%
e^{-2W}$ (ratio of the structure function at the second Bragg peak at
melting to its value at $T=0$) is about $60\%$. Using methods of \cite
{LiCorr}, one obtains for the 3D LLL GL model 
\begin{equation}
e^{-2W}=0.59.
\end{equation}

\subsection{Magnetization jump at melting}

The scaled magnetization is defined by: 
\begin{equation}
m\left( a_{T}\right) =-\frac{d}{da_{T}}f_{eff}\left( a_{T}\right)
\label{mscaled}
\end{equation}
At the melting point $a_{T}^{m}=-9.5$ the magnetization jump $\Delta M$
divided by the magnetization at the melting on the solid side is 
\begin{equation}
\frac{\Delta M}{M_{s}}=\frac{\Delta m}{m_{s}}=.018
\end{equation}
This is compared on the inset of Fig.2 with experimental results of fully
oxidized $YBa_{2}Cu_{3}O_{7}$ \cite{Nishizaki} (rhombs) and optimally doped
untwinned $YBa_{2}Cu_{3}O_{7-\delta }$ \cite{Welp} (stars). The agreement is
quite good.

\subsection{Specific heat jump at melting}

In addition to the delta function like spike at melting for specific heat
experiments shows also a specific heat jump. The theory allows to
quantitatively estimate it.

The specific heat of the vortex lattice is $C=-T\frac{\partial ^{2}}{%
\partial T^{2}}F_{eff}$.\ The scaled specific heat is defined as 
\begin{eqnarray}
c &=&\frac{C}{C_{mf}}=-\frac{16\beta _{A}}{9t^{2}}\left( \frac{b\omega }{%
4\pi \sqrt{2}}\right) ^{4/3}f(a_{T})  \nonumber \\
+ &&\frac{4\beta _{A}}{3t^{2}}\left( b-1-t\right) \left( \frac{b\omega }{%
4\pi \sqrt{2}}\right) ^{2/3}f^{^{\prime }}(a_{T})  \label{specifich} \\
&&-\frac{\beta _{A}}{9t^{2}}\left( 2-2b+t\right) ^{2}f^{^{^{\prime \prime
}}}(a_{T})  \nonumber
\end{eqnarray}
where $C_{mf}$ is the mean field specific heat of solid, $\frac{H_{c2}^{2}T}{%
4\pi \kappa ^{2}\beta _{A}T_{c}^{2}}$.

Thus the specific heat jump is: 
\begin{equation}
\Delta c=0.0075\left( \frac{2-2b+t}{t}\right) ^{2}-0.20Gi^{1/3}\left(
b-1-t\right) \left( \frac{b}{t^{2}}\right) ^{2/3}  \label{spjump}
\end{equation}
Using the parameters obtained fitting the melting line, we compare eq.(\ref
{spjump}) with the experimental result of ref. \cite{Schilling} in the inset
of Fig.2.\bigskip

\section{Higher Landau Levels contributions}

\subsection{Where is the LLL approximation really valid?}

Contributions of HLL are important phenomenologically in two regions of the
phase diagram. The first is at temperatures above the mean field $T_{c}$
inside the liquid phase.\bigskip\ The second is far below the melting point
deep inside the solid phase.\bigskip

Naively when ''distance from the mean field transition line'' is smaller
than the ''inter Landau level gap'', $1-t-b<2b$ one expects that higher
Landau harmonics can be neglected. More careful examination shows that a
weaker condition $1-t-b<12b$ should be used for a validity test of the LLL
approximation \cite{LiHLL} to calculate {\it the mean field contributions}
in vortex solid. Additional factor $6$ comes from the hexagonal symmetry of
the lattice since contributions of higher Landau levels (HLL) , \ first to
fifth HLL do not appear in perturbative calculation. How what about
contributions beyond mean field? The question has been studied by Lawrie 
\cite{Lawrie} in the framework of the Hartree - Fock (gaussian)
approximation in the vortex liquid. The result was that the region of
validity is limited. In this section we will incorporate the leading HLL
correction using gaussian approximation and then compare the theoretical
results with experimental magnetization curves.

\subsection{Gaussian Approximation in the liquid phase}

The free energy density is: 
\begin{equation}
F_{eff}=-\frac{\omega H_{c2}^{2}}{2\pi \kappa ^{2}V}ln\int D\psi D\overline{%
\psi }\exp \left( -f\right)
\end{equation}
where $f,V$ is the scaled full GL model eq.(\ref{GLs1}) and the scaled
volume. In gaussian approximation, $f$ \ is divided into an optimized
quadratic part $K$, and a ''small'' part $v.$ Then $K$ is chosen in such a
way that the gaussian energy is minimal. The gaussian energy is a rigorous
lower bound on energy. Due to translational symmetry of the vortex liquid an
arbitrary symmetric quadratic part $K\ $has only one variational parameter $%
\varepsilon ,$ 
\begin{equation}
K=\frac{1}{\omega }\left( \frac{1}{2}\left( |{\bf D}\psi |^{2}-b|\psi
|^{2}\right) +\frac{1}{2}|\partial _{z}\psi |^{2}+\varepsilon |\psi
|^{2}\right)
\end{equation}
The small perturbation is therefore: 
\begin{equation}
v=\frac{1}{\omega }\left[ \left( -a_{h}-\varepsilon \right) |\psi |^{2}+%
\frac{1}{2}|\psi |^{4}\right] .
\end{equation}
The gaussian energy consists of two parts. The first is the Trlog term: 
\begin{equation}
-\frac{\omega H_{c2}^{2}}{2\pi \kappa ^{2}}\log \left[ \int {\cal D}\psi
\exp (-K)\right] =\frac{\omega H_{c2}^{2}}{2\pi \kappa ^{2}}\frac{1}{\sqrt{2}%
\pi }b\sum_{n=0}^{\infty }\sqrt{nb+\varepsilon },
\end{equation}
The second is proportional to the expectation value of $v$ in a solvable
model defined by $K$

\begin{equation}
\frac{\omega H_{c2}^{2}}{2\pi \kappa ^{2}}\left\langle v\right\rangle =\frac{%
\omega H_{c2}^{2}}{2\pi \kappa ^{2}}\left[ \left( -a_{h}-\varepsilon \right) 
\frac{b}{2\sqrt{2}\pi }\sum_{n=0}^{\infty }\frac{1}{\sqrt{nb+\varepsilon }}%
+\omega \left( \frac{b}{2\sqrt{2}\pi }\sum_{n=0}^{\infty }\frac{1}{\sqrt{%
nb+\varepsilon }}\right) ^{2}\right] .
\end{equation}
Both are divergent in the ultraviolet. Introducing a UV momentum cutoff
which effectively limits the number of Landau levels to $N_{f}=\frac{\Lambda 
}{b}-1$, the Trlog term diverges as: 
\begin{equation}
\frac{1}{\sqrt{2}\pi }b\sum_{n=0}^{\infty }\sqrt{nb+\varepsilon }=\frac{1}{%
\sqrt{2}\pi }\left[ \frac{2}{3}\Lambda ^{3/2}+\left( \varepsilon -\frac{b}{2}%
\right) \Lambda ^{1/2}\right] +f_{tr\log }(\varepsilon ,b)  \label{regu}
\end{equation}
with the last term finite. The ''bubble'' integral diverges logarithmically: 
\begin{equation}
\frac{b}{2\sqrt{2}\pi }\sum_{n=0}^{\infty }\frac{1}{\sqrt{nb+\varepsilon }}=%
\frac{1}{\sqrt{2}\pi }\Lambda ^{1/2}+\frac{\partial }{\partial \varepsilon }%
f_{tr\log }(\varepsilon ,b)
\end{equation}
Substituting eq.(\ref{regu}) \ into the gaussian energy one obtains: $\ $(in
units of $\frac{\omega H_{c2}^{2}}{2\pi \kappa ^{2}}$) 
\begin{eqnarray}
&&\frac{1}{\sqrt{2}\pi }\frac{2}{3}\Lambda ^{3/2}+\omega (\frac{1}{\sqrt{2}%
\pi }\Lambda ^{1/2})^{2}+\left( -a_{h}-\frac{b}{2}\right) \frac{1}{\sqrt{2}%
\pi }\Lambda ^{1/2}-a_{h}\partial _{\varepsilon }f_{tr\log }(\varepsilon
,b)+2\omega \frac{1}{\sqrt{2}\pi }\Lambda ^{1/2}\partial _{\varepsilon
}f_{tr\log }(\varepsilon ,b) \\
&&-\varepsilon \partial _{\varepsilon }f_{tr\log }(\varepsilon ,b)+\omega
(\partial _{\varepsilon }f_{tr\log }(\varepsilon ,b))^{2}+f_{tr\log
}(\varepsilon ,b).  \nonumber
\end{eqnarray}

$T_{c}$ will be renormalized. We define $a_{h}=a_{h}^{r}+2\omega \frac{1}{%
\sqrt{2}\pi }\Lambda ^{1/2}$, the above energy becomes (in units of $\frac{%
\omega H_{c2}^{2}}{2\pi \kappa ^{2}}$): 
\begin{eqnarray}
&&\frac{1}{\sqrt{2}\pi }\frac{2}{3}\Lambda ^{3/2}-\omega (\frac{1}{\sqrt{2}%
\pi }\Lambda ^{1/2})^{2}+\left( -a_{h}^{r}-\frac{b}{2}\right) \frac{1}{\sqrt{%
2}\pi }\Lambda ^{1/2}-a_{h}^{r}\partial _{\varepsilon }f_{tr\log
}(\varepsilon ,b)  \label{reneq} \\
&&-\varepsilon \partial _{\varepsilon }f_{tr\log }(\varepsilon ,b)+\omega
(\partial _{\varepsilon }f_{tr\log }(\varepsilon ,b))^{2}+f_{tr\log
}(\varepsilon ,b).  \nonumber
\end{eqnarray}
Thus the temperature and vacuum energy will be renormalized. The first three
terms are divergent and linear in temperature, they will not contribute to
any physical quantities, like magnetization and specific heat. Minimizing
the energy eq.(\ref{reneq}), we get\ the gap equation: 
\begin{equation}
\varepsilon =-a_{h}^{r}-2\omega \partial _{\varepsilon }f_{tr\log
}(\varepsilon ,b)  \label{gapeq}
\end{equation}
Superscript ''r'' will be dropped later on. Function $f_{tr\log
}(\varepsilon ,b)$ can be written 
\begin{equation}
f_{tr\log }(\varepsilon ,b)=\frac{1}{\sqrt{2}\pi }b^{3/2}g\left( \frac{%
\varepsilon }{b}\right)
\end{equation}

where 
\begin{equation}
g\left( x\right) =\sum_{n=0}^{\infty }\left[ \sqrt{n+x}-\frac{2}{3}(x+n+%
\frac{1}{2})^{\frac{3}{2}}+\frac{2}{3}(x+n-\frac{1}{2})^{\frac{3}{2}}\right]
-\frac{2}{3}(x-\frac{1}{2})^{\frac{3}{2}}
\end{equation}
For the LLL model in the gaussian approximation, $g\left( x\right) =\sqrt{x} 
$. For Prange limit\cite{Prange} $Gi\rightarrow 0$, the free energy will be $%
\frac{\omega H_{c2}^{2}}{2\pi \kappa ^{2}}\frac{1}{\sqrt{2}\pi }%
b^{3/2}g\left( -\frac{a_{h}}{b}\right) $.

\bigskip

\subsection{Magnetization}

\bigskip When $\kappa $ is quite large and magnetization can be approximated
by 
\begin{equation}
M=-\frac{\partial }{\partial B}F_{eff}(B).  \label{magful}
\end{equation}

The HLL correction will be calculated as follows. We numerically solve the
gap equation (\ref{gapeq}) , and $F_{eff}(B)$ can be obtained. Then eq.(\ref
{magful}) is used to calculate the magnetization of the full GL model in
gaussian approximation. The HLL correction is thus the magnetization of the
full GL model in {\it gaussian approximation} minus the magnetization of the
LLL contribution in {\it gaussian approximation}. We compare the experiments
using following approximation. The LLL contribution (for example, see ref.(
\cite{LiPRL})) is  
\begin{equation}
M_{LLL}=\frac{H_{c2}}{4\pi \kappa ^{2}}\frac{a_{h}}{a_{T}}m\left(
a_{T}\right) ,  \label{magLLL}
\end{equation}
where $m\left( a_{T}\right) $ is defined in eq.(\ref{mscaled}) is taken
exactly (calculated by the BP method to highest order), while the
corrections due to HLL are taken in gaussian approximation. The comparison
of the theoretical predictions with the experiments for fully oxidized $%
YBa_{2}Cu_{3}O_{7}$\cite{Nishizaki}, is shown on Fig. 3. We use the
experimental asymmetry value $\gamma =4$ and values of $T_{c},$ $H_{c2}$ and 
$Gi$ from the fitting of the melting curve (see Fig.2 and Table 1). The
agreement is fair at high magnetic fields, while at low magnetic fields is
not good. We comment that \ the theory of the full GL model (higher Landau
levels included) beyond Gaussian approximation is required at low magnetic
fields.

\section{Summary}

The problem of calculating the fluctuations effects in the framework of the
Ginzburg - Landau approach is discussed and essentially solved in the LLL
limit. The melting line location is determined and magnetization and
specific heat jumps along it are calculated . The magnetization of liquid is
larger than that of solid by 1.8\% irrespective of the melting temperature,
while the specific heat jump is about 6\% and decreases slowly with
temperature. The LLL results for melting line, magnetization jump, specific
heat jump and the magnetization curves are in good agreement with
experiments and Monte Carlo simulations. The leading corrections due to
higher Landau levels have been also calculated and lead to agreement with
experimental results in the whole region in which the GL approach is
applicable.

\acknowledgments We are grateful to our colleagues T.-M. Uen, J.-Y. Lin for
numerous discussions and T. Nishizaki, A. Junod and M. Naughton for
providing details of experiments. The work was supported by NSC of Taiwan
grant NSC89-2112-M-009-039.

{\Huge Figure captions}


{\LARGE Fig. 1}

Free energy of solid (line) and liquid (dashed line) as function of the
scaled temperature. The solid line ends at a\ point (dot) indicating the
loss of metastability. Inset shows a tiny difference between liquid and
solid near the melting point at $a_{T}^{m}=-9.5$.

{\LARGE Fig. 2}

Comparison of the experimental melting line for fully oxidized $%
YBa_{2}Cu_{3}O_{7}$ \cite{Nishizaki} with our fitting. Inset on the right
shows the relative universal magnetization jump of $1.8\%$ (line) and
experimental results for fully oxidized $YBa_{2}Cu_{3}O_{7}$ \cite{Nishizaki}
(rhombs) and optimally doped untwinned $YBa_{2}Cu_{3}O_{7-\delta }$ \cite
{Welp} (stars). Inset on the left shows the relative nonuniversal specific
heat jump (line) and experimental results for optimally doped untwinned $%
YBa_{2}Cu_{3}O_{7-\delta }$ \cite{Schilling}.

{\LARGE Fig. 3}

Comparison of the theoretical magnetization curves (lines) of fully oxidized 
$YBa_{2}Cu_{3}O_{7}$ utilizing parameters obtained by fitting the melting
line on Fig.2 with torque magnetometry experimental results \cite{Nishizaki}
(dots). Arrows indicate melting points while at low magnetic field the
experimental data start from the point in which the magnetization is
reversible indicating low disorder.

\end{document}